\documentclass[twocolumn]{aastex62}
\submitjournal{ApJ}

\shorttitle{RS Oph: Expanding Bipolar X-ray Structure}
\shortauthors{Montez et al.}
\begin{document}

\title{Expanding Bipolar X-ray Structure After the 2006 Eruption of RS Oph}

%% \author[xxxx-xxxx-xxxx-xxxx]{Author Name}

\correspondingauthor{R. Montez Jr.}
\email{rodolfo.montez.jr@gmail.com}

\author[0000-0002-0786-7307]{R. Montez Jr.}
\affil{Center for Astrophysics $\vert$ Harvard \& Smithsonian, 60 Garden St., Cambridge, MA 01238, USA}

\author[0000-0002-2647-4373]{G. J. M. Luna}
\affil{CONICET-Universidad de Buenos Aires, Instituto de Astronom\'ia y F\'isica del Espacio (IAFE), Av. Inte. G\"uiraldes 2620, C1428ZAA, Buenos Aires, Argentina}
\affiliation{Universidad de Buenos Aires, Facultad de Ciencias Exactas y Naturales, Buenos Aires, Argentina.}
\affiliation{Universidad Nacional de Hurlingham, Av. Gdor. Vergara 2222, Villa Tesei, Buenos Aires, Argentina}

\author[0000-0001-8067-8732]{K. Mukai}
\affiliation{CRESST II and X-ray Astrophysics Laboratory, NASA Goddard Space Flight Center, Greenbelt, MD 20771, USA}
\affiliation{Department of Physics, University of Maryland, Baltimore County, 1000 Hilltop Circle, Baltimore, MD 21250, USA}

\author[0000-0003-2835-0304]{J. L. Sokoloski}
\affiliation{Columbia Astrophysics Lab 550 W120th St., 1027 Pupin Hall, MC 5247 Columbia University, New York, New York 10027, USA }

\author[0000-0002-3138-8250]{J. H. Kastner}
\affiliation{Rochester Institute of Technology, Rochester, NY}

\begin{abstract}
We report on the detection and analysis of extended X-ray emission by the {\it Chandra} X-ray Observatory stemming from the 2006 eruption of the recurrent novae RS Oph. 
The extended emission was detected 1254 and 1927 days after the start of the 2006 eruption and is consistent with a bipolar flow oriented in the east-west direction of the sky with opening angles of approximately $70^{\circ}$. 
The length of both lobes appeared to expand from $1\farcs3$ in 2009 to $2\farcs0$ in 2011, suggesting a projected expansion rate of $1.1\pm0.1 {\rm ~mas~day}^{-1}$ and an expansion velocity of $4600\ {\rm km~s}^{-1}\ (D/2.4\ {\rm kpc})$ in the plane of the sky. 
This expansion rate is consistent with previous estimates from optical and radio observations of material in a similar orientation. 
The X-ray emission does not show any evidence of cooling between 2009 and 2011, consistent with free expansion of the material. 
This discovery suggests that some mechanism collimates ejecta away from the equatorial plane, and that after that material passes through the red-giant wind, it expands freely into the cavity left by the 1985 eruption.  
We expect similar structures to arise from latest eruption and to expand into the cavity shaped by the 2006 eruption. 
\end{abstract}


\section{Introduction} \label{sec:intro}

RS Oph is a well-studied symbiotic recurrent nova (RN) experiencing outbursts separated by quiescent periods of between 9 and $\sim$21 years \citep{2010ApJS..187..275S}. 
On 2006 February 12.83 UT, RS Oph experienced its 6th recorded outburst \citep{2006IAUC.8671....2N}.  The 7th outburst was detected on 2021 August 8.93 UT (CBET 5013). 
The symbiotic system is comprised of a red giant \citep[M2IIIpe+;][]{2000AJ....119.1375F} and a massive white dwarf \citep[$M= 1.2-1.4~M_{\sun}$; e.g.,][]{2017ApJ...847...99M}.

The white dwarf (WD) accretes hydrogen-rich material from the red giant and experiences nova eruptions when a thermonuclear runaway occurs on the surface of the WD. 
Recurrent novae like RS Oph are potential progenitors of Type Ia supernova \citep[e.g.,][]{2017ApJ...847...99M}, as the WD masses appear to be approaching the Chandrasekhar limit through the accretion of mass from the companion star.  

When the donor is a red giant, as in RS Oph, the mass liberated in every nova eruption drives a blast wave and other shocks into the circumstellar environment.
In terms of theoretical expectations, numerical simulations suggest that the ejecta must be asymmetric \citep[e.g.,][]{2008A&A...484L...9W,2009A&A...493.1049O,2016MNRAS.457..822B}.
\citet{2016MNRAS.457..822B} simulated three nova eruptions, each following 18-years of mass transfer during quiescence before the nova eruption was injected. 
Based on the simulations, the interaction of the nova ejecta with the circumstellar material is expected to generate an equatorial ring and a larger-scale bipolar flow. 

After the 2006 eruption, bipolar structure was indeed detected on time scales from days to years at radio, infrared (IR), and optical wavelengths. 
Interferometric observations of Br$\gamma$ acquired 5.5 days after the outburst with the AMBER/VLTI indicated a slowly expanding ring-like structure and faster structure elongated in E-W direction \citep{2007A&A...464..119C}.
The radial velocities of the ring-like and E-W structures were 1800 km/s and 2500-3000 ${\rm km~s}^{-1}$, respectively.  
This pattern of faster moving material in the E-W direction and a slower moving ring-like component closer to the stellar system was also seen in radio observations taken between 14 and 93 days after outburst \citep{2006Natur.442..279O, 2008ApJ...688..559R, 2008ApJ...685L.137S}. 
The two components were also seen in narrowband optical imaging by the HST acquired hundreds of days (155 and 449 days) after the 2006 eruption \citep{2007ApJ...665L..63B,2009ApJ...703.1955R} indicating the presence of an expanding bipolar structure with the lobes oriented in the East-West direction.

In this article, we use multiple observations of RS Oph by the {\it Chandra} X-ray Observatory that followed the 2006 eruption to show that the E-W bipolar structure seen in radio and optical observations also dominated the X-ray remnant for at least five years after the 2006 eruption. The images all contain a bright compact source consistent with the location of the RS Oph binary system \citep{2011ApJ...737....7N} and resolved extended X-ray emission. Here we focus on the extended X-ray emission surrounding RS Oph. We find that the extended X-ray emitting material had a constant temperature indicative of internal shocks, and that between 2009 and 2011, it expanded freely at approximately 6000~km~s$^{-1} ({\rm d}/2.4~{\rm kpc})$. 

\section{Observations and Data Preparation} \label{sec:observations}

\subsection{Observations}

We analyzed grating-less, imaging observations of RS Oph acquired with the Advanced CCD Imaging Spectrometer (ACIS) back-illuminated S3 chip on the {\it Chandra} X-ray Observatory.
Four observations were performed in the period of 2007 and 2011 (see Table~\ref{tab:observations}). 
In 2011, the two observations of RS Oph were acquired within a period of two days. 
We studied each 2011 observation individually but merged them for our final analysis.
All observations were made using the 1/4 subarray of the S3 chip so as to decrease the frame time and mitigate the potential pile-up of X-ray photons. 
As a result, the observations are free of pile-up effects. 

All observations were reduced and analyzed with the latest software and calibration products.
We used the {\it Chandra} X-ray analysis software CIAO (version 4.12) with the calibration database (CALDB) version 4.9.2.1. 
All observations were processed using the pixel event repositioning (SER) algorithm described in \citet{2003ApJ...590..586L, 2004ApJ...610.1204L} and implemented within the \texttt{chandra\_repro} data reprocessing script. 

We note that there is a known artifact in the empirical Chandra point spread function (PSF) that has been present since late 1999 but was not discovered and reported until mid-2010 \citep{2010HEAD...11.4011J}. 
The PSF artifact is a ``hook-like'' asymmetry with an extent of $\lesssim1^{\prime\prime}$ and with a position angle fixed with respect to the spacecraft roll angle. 
The CIAO tool \texttt{make\_psf\_asymmetry\_region} can be used to identify the region that might be influenced by the PSF asymmetry. 
The presence of the artifact in HRC and ACIS imaging observations suggests the origin of the artifact lies with {\it Chandra}'s High Resolution Mirror Assembly (HRMA). 

\begin{deluxetable}{lccr}
\tablecaption{{\it Chandra} Imaging Observations of RS Oph\label{tab:observations}}
\tablecolumns{5}
\tablewidth{0pt}
\tablehead{
\colhead{ObsID} &
\colhead{Obs. Start} &
\colhead{$t_{\rm exp}$} & \colhead{$t_{\rm outburst}$\tablenotemark{a}} \\ 
\colhead{} & \colhead{(UTC)} & \colhead{(ks)} & \colhead{(days)} }
\startdata
7457 & 2007-08-04T06:11:26 & 91.0 & 537.43 \\ 
9952 & 2009-07-20T16:13:45 & 124.7 & 1253.85 \\
12404 & 2011-05-23T22:21:53 & 83.9 & 1926.10 \\ 
12403 & 2011-05-25T10:15:44 & 150.9 & 1927.60 \\ 
\enddata
\tablenotetext{a}{Days since the outburst, $t_{\rm outburst}$, is the time between the reported outburst on 2006 February 12.83 UT \citep{2006IAUC.8671....2N} and the {\it Chandra} observation start time.}
%\tablecomments{}
\end{deluxetable}

\begin{figure*}[ht!]
\plotone{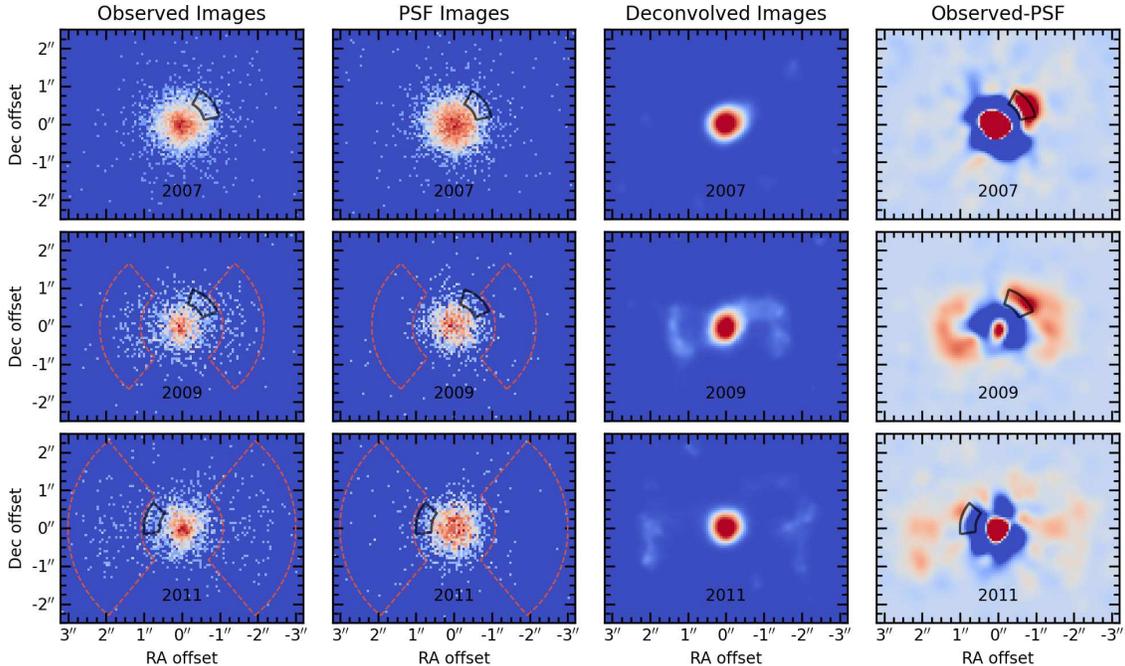}
\caption{Chandra observations of RS Oph. 
From left to right are the observed image, synthetic PSF image, deconvolved image, and observed minus PSF image. 
From top to bottom are the series of images for 2007, 2009, and 2011. 
Observed and PSF images were binned to 1/8 the native ACIS-S pixel size of $0\farcs492$.  
Deconvolved image was smoothed with a Gaussian kernal with FWHM$\sim 0\farcs15$. 
The PSF and observed images were smoothed with a Gaussian kernel with FWHM$\sim0\farcs4$ before subtracting. 
The black partial wedge-shaped regions indicate the approximate locations of the PSF asymmetry and the red, dashed-line wedge-shapes indicate spectral extraction regions are are shown on the observed and PSF images.  
\label{fig:analysis}}
\end{figure*}

\subsection{Synthetic PSF Generation}

A synthetic PSF is required for each observation to complete our analysis. 
Each PSF requires the source spectral characteristics, source location on the chip, and the observatory pointing information. 
The central point source spectra were extracted for each observation and fit using a combination of two absorbed thermal plasma models. 
We used Sherpa \citep{2001SPIE.4477...76F} for spectral fitting.
When fitting the source spectra, the goal was to accurately reproduce the photon energy distributions so that each synthetic PSF will be spectroscopically similar to each source.
The resulting spectroscopic models are provided as input into the PSF simulations. 

We used the {\it Chandra} Ray Tracer (ChaRT v2) to simulate bundles of rays from the spectroscopic models and pointing information of the observations.
ChaRT was used to simulate the rays passing through {\it Chandra} High Resolution Mirror Assembly (HRMA), then MARX \citep[version 5.3.3][]{2012SPIE.8443E..1AD} was used to project the rays onto the detector (ACIS-S3).
For the MARX projection we used source location on the detector and telescope pointing information from each observation. 
The parameters included the telescope roll information, detector configuration, subarray frame time, and the starting time of the observation. 
Setting the starting time (TStart) of MARX simulation, which was taken from the observation start time, is important to account for the contamination build-up on the ACIS optical blocking filters \citep{2018SPIE10699E..6BP}. 
As suggested by the MARX PSF analysis thread\footnote{\url{https://cxc.cfa.harvard.edu/ciao/why/aspectblur.html}}, we used an \texttt{AspectBlur} of $0\farcs25$ and included the \texttt{EDSER} subpixel adjustment option.
These analysis steps resulted in a series of synthetic PSF event lists, one for each observation. 

\subsection{Energy Filtering and Pixel Binning}

We filtered the source and PSF event lists to the 0.5-1.8 keV energy range. 
This energy range optimizes the signal in the extended emission while limiting increased background from the excluded energy ranges. 
The observed and PSF images were created with 1/8 binning of the native ACIS pixel size ($1/8 \times 0\farcs492$) and displayed in Figure~\ref{fig:analysis}. 
The analyses were also performed with 1/4, 1/3, and 1/2 binning of the native ACIS pixel size with similar results.
The location of the PSF asymmetry region is indicated for each observation (see Figure~\ref{fig:analysis}). 
Note that currently the HRMA model used by ChaRT does not include the PSF asymmetry.

\section{Analysis} \label{sec:analysis} 

Our analysis began with subtraction of a synthetic point spread function (PSF) from the observed image and image restoration through deconvolution of the observed image by the synthetic PSF.
We then used the restored image to derive profiles of the extended emission. 
The profiles were used to estimate the properties of the extended X-ray emission. 

\subsection{PSF Subtraction}

PSF subtraction was performed by smoothing the observed and PSF images with a Gaussian kernel with a full width at half maximum (FWHM) of $\sim0\farcs4$ then subtracting the PSF image from the observed image. 
The rightmost column of images in Figure~\ref{fig:analysis} show the results of the PSF subtraction. 

\subsection{Image Restoration}

\begin{figure}
\plotone{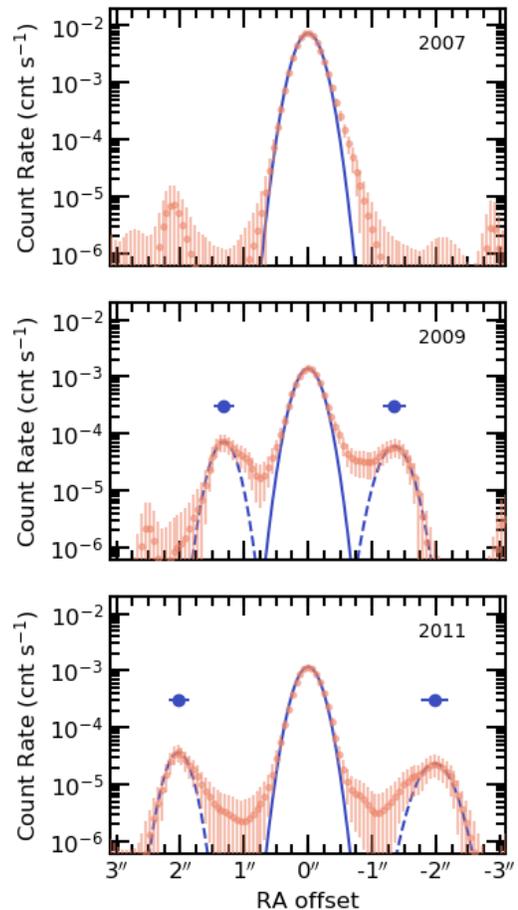}
\caption{Profiles derived from PSF-deconvolution images. 
The profiles are integrated along the Declination axis and divided by the total exposure time. 
Poisson statistics are assumed to determine the error bars at each point. 
The solid line indicates a Gaussian with FWHM$=0\farcs4$, which matches the stopping criteria for the image restoration. 
In the 2009 and 2011 observations, the properties of the extended X-ray emission are derived from the dashed curves which are the Gaussian functions summarized in Table~\ref{tab:extendedprops}. 
The East-West extent of the extended emission is summarized by the larger blue circular points that are plotted above the profiles. \label{fig:profiles}}
\end{figure}

Image restoration is performed on the observed image with the Richardson-Lucy maximum-likelihood deconvolution algorithm \citep{1972JOSA...62...55R,1974AJ.....79..745L}. 
The stopping criteria we adopted for image restoration were based on the characteristics of the bright central point sources in each observation. 
Specifically, we iterated the deconvolution algorithm until the central source reached a FWHM of $\sim0\farcs4$ (see Figure~\ref{fig:profiles} and next section). 
There was a range in the number of iterations (14-18) that reached the FWHM criteria, so we adopted 16 iterations. 
The algorithm can continue for many more iterations, but given the complexity of the extended emission, we opted for fewer iterations \citep{2012A&A...539A.133P}. 
The restored image was smoothed with a Gaussian kernel with a FWHM$\sim0\farcs15$ and presented in Figure~\ref{fig:analysis}.

\subsection{East-West Profiles}

Profiles of the extended emission are formed by integrating the restored image along the vertical North-South Declination axis in Figure~\ref{fig:analysis}. 
Since the source is oriented in the East-West direction, integrating along the Declination axis provides the maximum extent of the extended X-ray emission. 
The range of the integration on the Declination axis was limited to omit spurious background signals in the restored images. 
For each point along the Right Ascension direction we calculated the total counts and errors.
We assumed Poisson statistics for the error calculation and then divided all our measurements by the total exposure time to convert to count rates.
The resulting profiles are presented in Figure~\ref{fig:profiles} and are overlaid with Gaussians with FWHM$=\sim0\farcs4$ and normalized to the central sources. 
This Gaussian represents the stopping criteria described in the previous section. 

At the time of the publication of the 2007 observation \citep{2009ApJ...707.1168L}, the {\it Chandra} PSF anomaly was not widely known nor reported \citep{2010HEAD...11.4011J}. 
Based on this updated analysis, we determine that the jet-like feature reported in \citet{2009ApJ...707.1168L} is influenced by and likely due to the PSF anomaly.
For the remainder of this study, we only consider extended emission from the 2009 and 2011 observations. 

\subsection{Extended Emission Spatial Analysis}\label{sec:profiles}

To determine the extent of the extended X-ray emission we used Gaussian functions to match the left and right extended profiles. 
The normalization, $A$, centroid, $\mu$, and width, $\sigma$ of the various Gaussian kernels were adjusted until they approximated the profile distributions (see Figure~\ref{fig:profiles}).  
The resulting Gaussian parameters are then used to estimate the physical properties of the extended X-ray emission. 
The Gaussian centroids are used to estimate the extent of the extended X-ray emission in both the Eastern and the Western directions and we adopt the $\sigma$ values as an error estimate on the extent (see Figure~\ref{fig:profiles}).
To arrive at the combined value for each year, we averaged the extent in the Eastern and Western directions and add their $\sigma$ values in quadrature. 
$A$ in counts and $\sigma$ in pixel values are used to estimate the total counts and count rates of the extended emission ($\sqrt{2\pi} \sigma A$).
These estimates are listed in Table~\ref{tab:extendedprops}. 

The total count rate of the extended emission in 2009 and 2011 was 0.94$\pm$0.09 and  0.45$\pm$0.04 cnt ks$^{-1}$, respectively (see Table~\ref{tab:extendedprops}). 
Based on the effective area curves from the 2009 and 2011 observations, the total effective area over the 0.5-1.5 keV energy range has decreased by $\sim$8\% due to build up of the contaminant on the ACIS-S optical blocking filter. 
Hence, most of the $\sim$50\% reduction in the count rate was due to the extended emission fading. 

\begin{deluxetable}{lccccc}
\tablecaption{Extended Emission Properties\label{tab:extendedprops}}
\tablecolumns{5}
\tablenum{2}
\tablewidth{0pt}
\tablehead{
\colhead{Year} & \colhead{$A$} & \colhead{$|\mu|$} & \colhead{$\sigma$} & \colhead{Counts} & \colhead{Count Rate} \\ 
\colhead{} & \colhead{(cnt)} & \colhead{($^{\prime\prime}$)} & \colhead{($^{\prime\prime}$)} & \colhead{(cnt)} & \colhead{(cnt ks$^{-1}$)}  
}
\startdata
\sidehead{\it Eastern Direction:}
2009 & 9.0 & 1.32 & 0.16 & 59 & 0.47$\pm$0.06 \\
2011 & 8.6 & 2.03 &  0.16 & 56& 0.24$\pm$0.03 \\
\sidehead{\it Western Direction:} 
2009 & 7.5 & 1.35 & 0.19 & 58 & 0.47$\pm$0.06 \\
2011 & 5.5 & 1.98 & 0.22 & 49 & 0.21$\pm$0.03 \\
\sidehead{\it Combined:}
2009 &  & 1.34 & 0.25 & 117 & 0.94$\pm$0.09 \\ 
2011 &   & 2.01 & 0.27 & 105 & 0.45$\pm$0.04 \\
\enddata
\tablecomments{The extended emission properties are determined from Figure~\ref{fig:profiles} as described in \S\ref{sec:profiles}. Table columns include the normalization ($A$), centroid ($\mu$), and width ($\sigma$) of the Gaussian functions shown in Figure~\ref{fig:profiles}. 
The value of $\sigma$ must be converted to image pixels before calculating the total counts using: ${\rm Counts} = \sqrt{2\pi} \sigma_{\rm pixels} A$. Count Rates are calculated using the exposure times in Table~\ref{tab:observations}. Poisson statistics on the Counts are used to estimate the error on the Count Rates.  The combined value of $\mu$ is the average extension of the extended X-ray emission in the East-West direction. The combined values of $\sigma$ was determined by adding the individual $\sigma$ values for each year in quadrature. Adding the Counts for each year gives the combined Counts values and Count Rates.}
\end{deluxetable}

\begin{figure}
\plotone{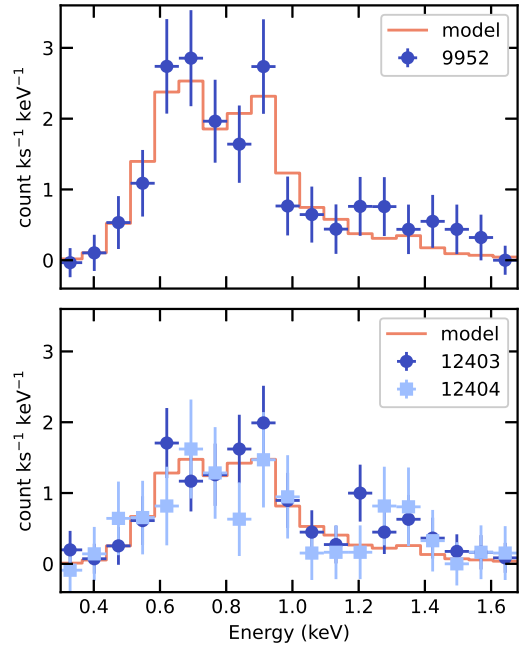}
\caption{X-ray spectra of the extended emission from RS Oph. {\it Top panel} depicts the spectrum from 2009 (ObsID 9952) and {\it bottom panel} depicts the spectrum from 2011 (ObsID 12403 \& 12404). In each panel, the best-fit model is shown by the solid red line. Other than changes due to the evolving effective area of the ACIS detector, the spectral shape did not change significantly between 2009 and 2011.
\label{fig:spectra}}
\end{figure}

\subsection{Extended Emission Spectral Analysis}

We extracted spectra of the extended emission from the 2009 and 2011 original observations. 
We limited our extraction to angles consistent with the extended emission seen in the restored image and used the synthetic PSF to reduce potential contamination from the bright central source. 
In Figure~\ref{fig:analysis}, the regions used for spectral extraction are overlaid on the observed images.
The Eastern and Western extended emission were extracted as a single spectrum for each observation (ObsIDs 9952, 12403, and 12404). 
Each spectrum was extracted with the CIAO \texttt{specextract} tool. 
The background for the energy range considered was negligible. 

We modeled the spectra with Sherpa using an absorbed thermal plasma model (\texttt{tbabs}$\times$\texttt{apec}). 
The two observations from 2011 were fit simultaneously with all model parameters tied together. 
Initially, we left the column density ($N_{\rm H}$) as a free parameter but found it difficult to constrain both the column density and the plasma model normalization. 
Instead, we adopted the $E_{\rm B-V} = 0.73 \pm 0.10\ {\rm mag}$ from \citet{1987Ap&SS.130..243S} and $R_V=3.1$ to set $N_{\rm H}$ to $5\times10^{21}\ {\rm cm}^{-2}$ \citep[also consistent with recent measurements by][]{2018MNRAS.480.1363Z}.
The adopted $N_{\rm H}$ value is similar to those values derived when $N_{\rm H}$ is left as a free parameter for the 2009 ($N_{\rm H}\sim4\times10^{21}\ {\rm cm}^{-2}$) and 2011 ($\sim6\times10^{21}\ {\rm cm}^{-2}$) spectra. 
The resulting best-fit parameters and derived fluxes (observed and intrinsic) are provided in Table~\ref{tab:spectralfits}. 
The spectra and best-fit models are presented in Figure~\ref{fig:spectra}. 

\begin{deluxetable}{lccccc}
\tablecaption{Extended Emission Spectral Fits\label{tab:spectralfits}}
\tablecolumns{3}
\tablenum{3}
\tablewidth{0pt}
\tablehead{
\colhead{Parameter} & \colhead{2009} & \colhead{2011} }
\startdata
$N_{\rm H}\ (10^{21}\ {\rm cm}^{2})$\tablenotemark{a} & 5 & 5 \\ 
$T_{\rm X}\ ({\rm MK})$ & $2.0_{-0.2}^{+0.2}$ & $2.1_{-0.2}^{+0.4}$ \\
$\eta\ (10^{-5}\ {\rm cm^{-5}})$\tablenotemark{b} & $10.2_{-3.9}^{+6.1}$ & $5.5_{-2.6}^{+2.6}$ \\
$F_{\rm X, obs}\ (10^{-15}\ {\rm erg\ cm}^{-2}\ {\rm s}^{-1})$ & $5.5_{-1.6}^{+1.6}$ & $3.5_{-0.9}^{+1.0}$ \\
$F_{\rm X}\ (10^{-14}\ {\rm erg\ cm}^{-2}\ {\rm s}^{-1})$ & $11.5_{-3.0}^{+2.6}$ & $6.3_{-2.9}^{+3.0}$ \\
\sidehead{\it Distance-dependent Properties\tablenotemark{c}:}
$L_{\rm X}\ (10^{31}\ {\rm erg}\ {\rm s}^{-1})$ & $7.9_{-3.3}^{+5.1}$ & $4.3_{-2.2}^{+2.3}$ \\
$EM\ (10^{54}\ {\rm cm}^{-3})$ & $7.1_{-2.9}^{+4.5}$ & $3.8_{-1.9}^{+2.1}$ \\
$V\ (10^{49}\ {\rm cm}^{3})$ & $2.1_{-1.4}^{+1.4}$ & $7.1_{-3.9}^{+3.9}$ \\
$n_{e}\ ({\rm cm}^{-3})$ & $630_{-250}^{+300}$ & $250_{-90}^{+110}$ \\
$M_{\rm sh}\ (10^{-5}\ M_{\odot})$ & $0.9_{-0.7}^{+0.8}$ & $1.3_{-0.8}^{+0.9}$ \\
\hline
\enddata
\tablecomments{The quoted errors for each parameter are for the 90\% confidence range.}
\tablenotetext{a}{The column density ($N_{\rm H}$) was frozen to $5\times10^{21} {\rm ~cm}^{-2}$ based on $E_{\rm B-V}=0.73 {\rm ~mag}$.}
\tablenotetext{b}{The $\eta$ parameter is the model normalization value.}
 \tablenotetext{c}{Formulae for the distance-dependent parameters are provided in \S\ref{sec:discussion}.}
\end{deluxetable}

\section{Results} 

Our analysis reveals the presence of extended X-ray emission from RS Oph in the imaging observations by {\it Chandra} in 2009 and 2011 (see Figure~\ref{fig:analysis}). 
The images all contain a bright compact source consistent with the location of the RS Oph binary system. 
To the east and west of the compact source, extended emission appears in the observed images from 2009 and 2011 but no such emission is apparent in the image from 2007 (top row in Figure~\ref{fig:analysis}). 
The extended X-ray emission in the 2009 and 2011 observations was detected in the 0.4 to 1.6 keV energy range. 
After subtracting a smoothed synthetic PSF from the smoothed observed image, the presence of the extended X-ray emission is evident along with the known PSF asymmetry (rightmost column of images in Figure~\ref{fig:analysis}). 
When we performed our analysis, the ChaRT simulation of {\it Chandra}'s HRMA does not account for the PSF asymmetry, thus leading to an enhancement in the PSF-subtracted images where the asymmetry is expected. 

We restored the observed images by deconvolution with the synthetic PSFs to produce high definition images of the extended X-ray emission (third column of images in Figure~\ref{fig:analysis}). 
The restored images are integrated along the Declination axis to derive profiles of the extended X-ray emission in the East-West directions (see Figure~\ref{fig:profiles}). 
Based on the images and profiles, any extended X-ray emission in the 2007 observation can be attributed entirely to the PSF asymmetry. 
On the other hand, although some of the detected extended X-ray emission near the compact core in 2009 and 2011 can be attributed to the PSF asymmetry, we find extended X-ray emission beyond $\sim1^{\prime\prime}$ that is astrophysical in origin in both of those epochs.

The extended X-ray emission in the 2009 and 2011 observations is evident in the emission profiles (Figure~\ref{fig:profiles}). 
We used these profiles to derive the properties of the extended emission. 
The extent of the extended X-ray emission grew from 2009 to 2011.
The average distances between the central source and the peak of the features in Figure~\ref{fig:profiles} on the sky in 2009 and 2011 was $\sim1\farcs3$ and $\sim2\farcs0$, respectively. 
Assuming the extended emission originates from the outburst in 2006, these sizes are consistent with an expansion rate of just over $1\ {\rm mas~day}^{-1}$.
From our analysis of the profiles, we estimate that the count rate of the extended emission decreased by $\sim$50\% between the 2009 observation to the 2011 observation with $\sim10\%$ of that reduction due to a loss in instrumental sensitivity. 
Based on spectral fitting, the extended X-ray emission during both epochs was consistent with a plasma temperature of $\sim2~{\rm MK}$. 
The model-derived intrinsic flux, $F_{\rm X}$, of the source decreased by $35\pm13\%$, which is consistent with the decrease in the count rate after accounting for the loss of instrumental sensitivity. 

The opening angle, $\phi$, of the extended emission was roughly $70^{\circ}$ in both the eastern and western directions based on Figure~\ref{fig:analysis}. 
Arc-like structure was evident in the 2009 observation but was less coherent in 2011.  
Given the noise in the images, it is not possible to determine whether this apparent evolution in the spatial distribution was the result of actual physical changes.

\section{Discussion} \label{sec:discussion} 

Extended X-ray emission from RS Oph is consistent with plasma emission expanding and fading at a constant temperature. 
The orientation of the extended X-ray emission is consistent with post-eruption structures tracked in radio imaging \citep{2006Natur.442..279O, 2008ApJ...688..559R,2008ApJ...685L.137S} and narrowband optical imaging \citep{2007ApJ...665L..63B,2009ApJ...703.1955R}. 

Our results provide a rare, though not unique, glimpse into the shocked ejecta from novae. While similar features have been detected in DQ Her and several other systems \citep[see][and references therein]{2020MNRAS.495.4372T}, those observations were taken decades after the eruptions.  
In addition to the much shorter timescales (3--5 years after eruption), the key difference in the case of RS Oph is that we can take advantage of the wealth of multi-wavelength data obtained during the 2006 eruption.

\subsection{Expansion Rate of Extended Emission}
\begin{figure*}
\plotone{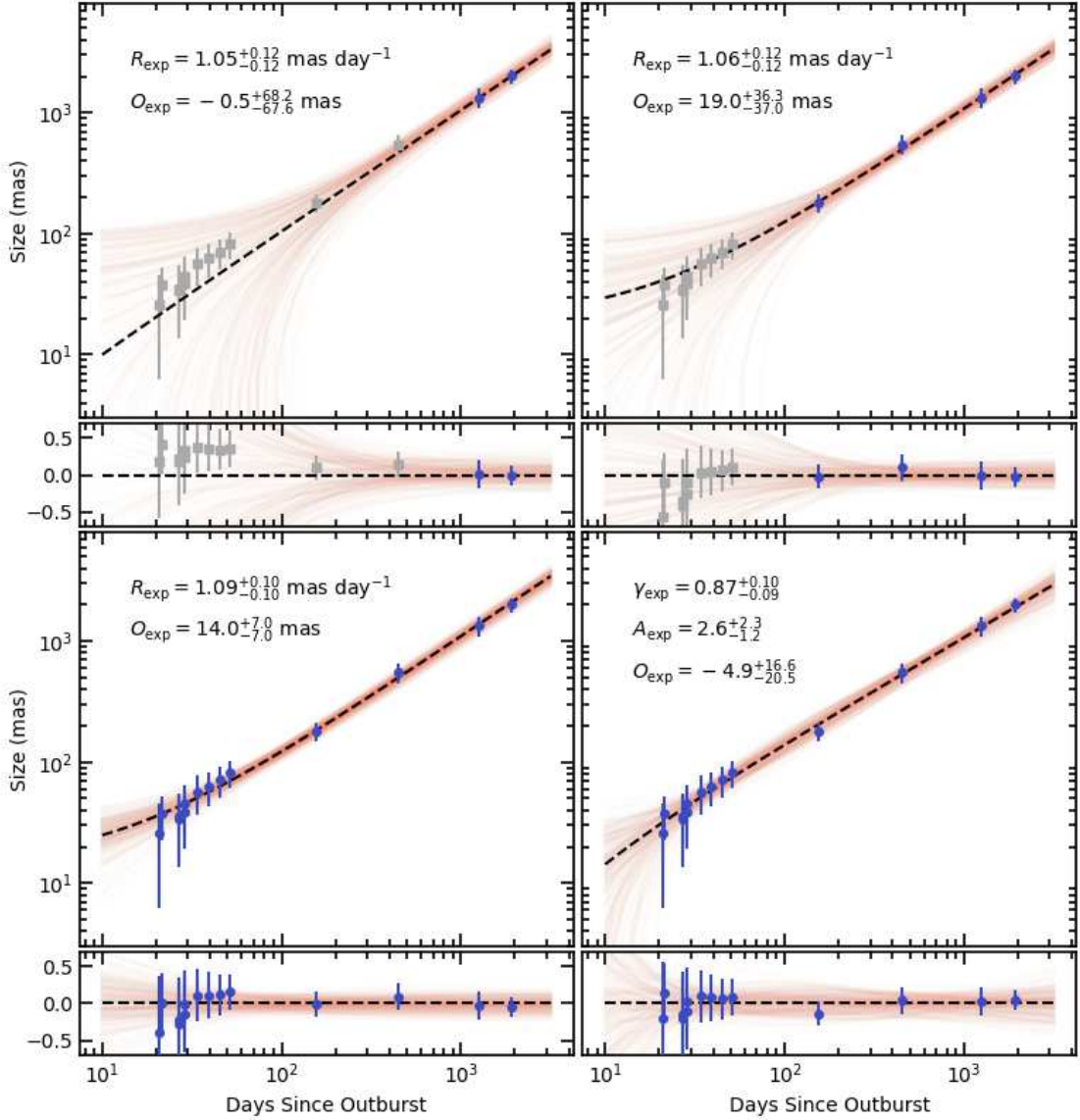}
\caption{Expansion of the size of East and West ejecta from RS Oph since the last outburst. Blue data (circles) are included in the MCMC analysis while the grey data (squares) are ignored. Panels depict the analysis results as one increases the observations included in the analysis. A random selection of 200 MCMC samples (out of 70,000) are shown in red while the black dashed line depicts the best-fit model based on the median values of posterior distributions of the model parameters, these values and their 68\% confidence range are included in each panel.  
The smaller panels show the residuals normalized by the size. 
All panels depict a linear expansion model, except the lower right panel, which depicts a powerlaw expansion model. 
\label{fig:expansion}}
\end{figure*}

Comparing the extent of the X-ray emission with multi-wavelength observations taken prior to our X-ray observations shows that the the X-ray structures were roughly consistent with an extrapolation of the radio and optical bi-polar flows. 
Early radio observations at day 21.5 to day 51 since the 2006 eruption were conducted by the European VLBI Network and the Very Long Baseline Array from 1.7-5 GHz and show extended emission from a ring-like structure and expanding emission in the eastern direction \citep{2006Natur.442..279O, 2008ApJ...688..559R,2008ApJ...685L.137S}. 
We used the images presented in these sources to estimate the angular size of the radio emission in the Eastern direction at each epoch and conservative uncertainty estimates based on the reported beam sizes, the width of the extended emission, and our estimate of the location of the binary system.
\citet{2007ApJ...665L..63B} and \citet{2009ApJ...703.1955R} reported a sequence of optical images 155 and 449 days after the 2006 eruption, respectively, taken with the Hubble Space Telescope that detected narrowband [OIII] 
$\lambda$5007 emission from a bipolar structure with the lobes oriented East-West direction with an inclination of $39^{+1}_{-10}\deg$. 
The lobes appeared to linearly expand in the East-West direction with no evidence of deceleration and an expansion rate of $1.2\pm0.1 {\rm ~mas~day}^{-1}$ \citep{2009ApJ...703.1955R}. 
We take their reported measurements of the full East-West angular size and divide by two to arrive at the angular size from the binary system. 
In Figure~\ref{fig:expansion}, we show collected extension angular size estimates for all epochs of the radio, optical, and our X-ray observations. 

We determined the expansion rate, $R_{\rm exp}$, of the extended emission assuming linear expansion from an origin, $O_{\rm exp}$, and estimating the model parameters with the \texttt{emcee} package \citep{2013PASP..125..306F}, an affine-invariant ensemble sampler for Markov chain Monte Carlo (MCMC). 
We ran 100,000 samples with uniform priors and disregarded the initial 30,000, leaving 70,000 samples from which we derived the posterior distributions of the model parameters. 
We used the median value of the posterior distribution as the best-fit value and the 16-th and 84-th percentile as the lower and upper error range (68\% confidence range, or a 1-$\sigma$ error range).  
We performed a series of three fits: (1)  only the X-ray data points (upper panel of Figure~\ref{fig:expansion}), (2) the X-ray and optical data points (central panel), and, (3) the X-ray, optical, and radio data points (lower panel). 
All linear expansion rates are consistent with $R_{\rm exp} = 1.1\pm0.1 {\rm ~mas~day}^{-1}$ and the $O_{\rm exp}$ is consistent with zero at least at the $2-\sigma$ level.
The constraints on $R_{\rm exp}$ and $O_{\rm exp}$ improve as we include more multiwavelength measurements.

Noticing that the best-fit $R_{\rm exp}$ values decrease from fit (1) to fit (3) --- but are consistent within their respective errorbars --- we also considered a powerlaw model for the multiwavelength expansion of the form $\mu = A_{\rm exp} t^{\gamma_{\rm exp}} + O_{\rm exp}$, where $\mu$ is the size of the extended emission, $A_{\rm exp}$ is the normalization, $\gamma_{\rm exp}$ is the powerlaw, and $O_{\rm exp}$ is the origin. 
The origin component is necessary in order to accurately compare this model to the linear expansion model since we do not assume $O_{\rm exp} = 0$. 
Performing a similar MCMC analysis with uniform priors on the powerlaw model parameters results in a solution that suggests some deceleration ($\gamma_{\rm exp}<1$) but only marginally at the 1-$\sigma$ level (see Figure~\ref{fig:expansion}). 
However, we can make a model comparison using the Bayes factors that result from integrating the multidimensional posterior distributions for each model. 
The ratio of the Bayes factors gives the odds of favoring one model over another, where odds ratios $>10$ indicate strong evidence for one model over another. 
We find the linear model is favored by an odds ratio of 1.6, suggesting the two models cannot be distinguished given the data. 
We adopt the linear expansion model. 

All the measurements are thus consistent with the linear expansion of the material in the East-West direction and earlier reports of expansion rates \citep{2006Natur.442..279O, 2008ApJ...688..559R,2008ApJ...685L.137S, 2009ApJ...703.1955R}. 
Incidentally, the best-fit $R_{\rm exp}$ value suggests that at the time of the 2007 {\it Chandra} observation, the extended emission would have measured $\sim0\farcs6\pm0\farcs3$, making its detection difficult. 
Although the emitting material and processes can be distinct in each observed spectral range, the material in the East-West bipolar flows all appears to be driven by an underlying process associated with the 2006 eruption. 
Furthermore, based on this analysis, there is no strong evidence that the bipolar flows are decelerating, suggesting that the East-West material has been freely expanding since or shortly after the 2006 eruption.

\subsection{Distance to RS Oph and Expansion Velocity} 

A distance is required to turn our multiwavelength expansion rate ($R_{\rm exp} = 1.1\pm{0.1} {\rm ~mas~day}^{-1}$) into an expansion velocity. 
The distance to RS~Oph is a subject of debate \citep{2008ASPC..401...52B,2009ApJ...697..721S} with values ranging from $<$1~kpc to $>5$~kpc. 
Gaia (early Data Release 3) parallax measurements of RS Oph, suggest a distance of $2.4^{+0.3}_{-0.2}$~kpc  \citep{2021AJ....161..147B}.
This measurement has a low Renormalized Unit Weight Error (RUWE$=1.29$) but a non-negligible excess astrometric noise ($\epsilon_{i} = 0.13$~mas) in Gaia eDR3 \citep{2016A&A...595A...1G,2021A&A...649A...1G}.
RUWE has been used in the literature to calculate astrometric wobble possibly induced by binary motions and $\epsilon_{i}$ has been found to track linearly with this astrometric wobble \citep{2020MNRAS.496.1922B}.
Adding $\epsilon_{i}$ in quadrature to the parallax error results in an increase in the error on the parallax measurement from $\sim 6\%$ to $\sim35\%$.
For the purpose of this article, we adopt the Gaia distance of $2.4^{+0.3}_{-0.2}$~kpc. 
For the remainder of the discussion, we use this adopted distance and provided distance-scaled formulae for our calculations. 

Using this adopted distance, we calculate the projected expansion velocity
\begin{eqnarray}
v_{\rm exp, proj} = 4600\pm700\ {\rm km~s}^{-1} \left(\frac{R_{\rm exp}}{1.1 {\rm ~mas~day}}\right) \nonumber \\ \times \left(\frac{D}{2.4 {\rm ~kpc}}\right),
\end{eqnarray}
where $R_{\rm exp}$ is the multiwavelength expansion rate and we have taken a conservative symmetric error on the Gaia distance. 
\citet{2009A&A...497..815B} studied optical and near-infrared spectroscopic observations the RS Oph system and determined a narrow orbital inclination of 49-53$^{\circ}$. Assuming the expanding material is perpendicular to the orbital plane, an inclination of 51$^{\circ}$ results in a deprojected velocity of $5900\pm900\, {\rm km~s}^{-1}$

Early IR observations shortly after the 2006 eruption show constant expansion for $\sim 4~d$ with an expansion velocity $\sim 3000~{\rm km~s}^{-1}$ before deceleration \citep{2006ApJ...653L.141D}, likely as the blast wave encountered denser equatorial material. 
\cite{2007A&A...464..119C}, $5.5~d$ after the 2006 eruption, identified a slower moving equatorial ring with radial velocity $\leq1800 {\rm ~km~s}^{-1}$ and a faster moving East-West bipolar flow with radial velocity of approximately $2500-3000 {\rm ~km~s}^{-1}$. 
Emission line profiles from spectroscopic X-ray observations 13.5 days after the eruption suggest the blast wave motion was primarily in the plane of the sky with an expansion velocity at that time of $2400\pm400 {\rm ~km~s}^{-1}$ \citep{2009ApJ...691..418D}. However, by Day 111, the blueshifts of the X-ray lines had decreased markedly, with the centroids of many lines being consistent with laboratory wavelengths \citep{2008ApJ...673.1067N}. These X-ray observations were also consistent with the picture of a sharply decelerating equatorial flow with a fast bipolar flow.

The temperature of the extended X-ray emission ($\sim$2 MK), assuming strong-shock conditions, implies a shock velocity of order $400 {\rm ~km~s}^{-1}$, much lower than the bulk velocity measured by the angular expansion of the X-ray emitting regions. 
Because the shock velocity is the velocity of the shock front relative to unshocked matter, there is no contradiction here. 
We suggest that the X-rays are from reverse shocks driven into the bipolar flow, which were carried outward by the momentum of that flow.

We suggest that the bipolar outflow from the 2006 eruption is now expanding in a cavity cleared out by the previous (1985) eruption of RS Oph. 
It is plausible that, in the polar directions, the outflows do not slow down significantly until they sweep up its own mass in ISM, which could take centuries. 
Until then, each eruption creates its own cavity, and only a fraction of the cavity volume is filled by the slow red giant wind. 
Given the typical wind velocity of red giant wind of a few 10s of ${\rm km~s}^{-1}$, or of order 1\% of the outflow velocity, the 2006 ejecta will break out of the post-1985 red giant wind region in a few tens of days. 
If the bipolar outflow was shocked during these early days, and is now expanding freely into the cavity, that could explain the linear expansion.

\subsection{Spectral Evolution of Extended Emission}

The spectral fits of the extended X-ray emission suggests no change in the temperature and marginal evidence for fading emission. 
Assuming free-free emission is the dominate cooling process, then the time scale for the gas to radiate its thermal energy is
\begin{equation}
t_{\rm cool} = 6\times10^{4} {\rm ~yr} \left(\frac{T_{e}}{10^6 {\rm ~K}}\right)^{1/2}  \left(\frac{n_{e}}{100 {\rm ~cm}^{-3}}\right)^{-1},
\end{equation}
where $T_{e}$ is the electron temperature, $n_{e}$ is the electron number density \citep{2002apa..book.....F}. 
Radiative cooling by emission lines, 
provides a more efficient cooling at temperatures of $\sim 2\times10^{6} {\rm ~K}$, decreasing the cooling time scale by a factor of $\sim 30$ times \citep{2015A&A...578A..15L}. 
Observable cooling on the time scale of these observations since eruption ($\sim 3-6 {\rm ~yr}$) would require $n_{\rm e} \gtrsim 10^{6} {\rm ~cm}^{-3}$. 

The density of the emitting material is related to the model normalization, $\eta$, by the emission measure, $EM = \int n_e n_{\rm H} dV$, and the formula: $\eta = \frac{10^{-14}}{4\pi D^2} EM$, where $D$ in the distance cm, $V$, is the volume in cm$^{3}$, and $n_{e}$ and $n_{\rm H}$ are the electron and hydrogen number densities, respectively, in cm$^{-3}$.
The emission measure is given by, 
\begin{equation}\label{eq:em}
EM = 6.9\times 10^{53} {\rm ~cm}^{-3} \left(\frac{\eta}{10^{-5} {\rm ~cm}^{-5}}\right)\left(\frac{D}{2.4\ {\rm kpc}}\right)^{2}.
\end{equation}
To estimate the volume, we assume the emission arises from thin shells limited to spherical sectors to the East and West. 
The volume of each shell can be approximated as the shell thickness, $l_{\rm sh}$, multiplied by the shell surface area, $2\pi r_{\rm sh}^{2} (1 - \cos(\phi/2))$, where $r_{\rm sh}$ is the size defined as the distance from RS Oph to the shell of extended emission and $\phi$ is the opening angle estimated from the position angles that encompass the extended emission in the deconvolved images ($\phi\sim70^{\circ}$).  
Table~\ref{tab:extendedprops} lists $\mu$, the angular distance on the sky from RS~Oph to the extended emission, from which we obtain,
\begin{equation}\label{eq:rshell}
r_{\rm sh} = 7 \times 10^{16} {\rm ~cm}\ \left(\frac{\mu}{2^{\prime\prime}}\right) \left(\frac{D}{2.4 {\rm ~kpc}}\right),
\end{equation}
where $D$ is the physical distance to RS~Oph.
For a strong shock with a compression ratio of 4, we take $l_{\rm sh}\approx r_{\rm sh}/12$. 
Then the total volume (adding the volume of the Eastern and Western shells together) is given by,
\begin{equation}\label{eq:volume}
V \approx 7\times10^{49} {\rm ~cm}^{3} \left(\frac{\mu}{2^{\prime\prime}}\right)^{3} \left(\frac{D}{2.4 {\rm ~kpc}}\right)^{3}.
\end{equation}
Assuming a constant density in the shell and solar abundances ($n_{\rm e} = 0.8 n_{\rm H}$),  
\begin{equation}\label{eq:density}
n_{\rm e} \approx 95 {\rm ~cm}^{-3} \left(\frac{EM}{6.9\times10^{53} {\rm ~cm}^{-3}}\right)^{1/2} \left(\frac{V}{7\times10^{49} {\rm ~cm}^{3}}\right)^{-1/2}.
\end{equation}
It should be noted that the constant density in the shell is unlikely to reflect the actual conditions.  Furthermore, by using the approximation $l_{\rm sh}\approx r_{\rm sh}/12$, we assume that the total swept mass is equal to the total shocked mass, again unlikely to reflect actual conditions. These assumptions introduce larger uncertainty than reflected in our calculations (see Table~\ref{tab:spectralfits}). 

Using Equations~\ref{eq:em}-\ref{eq:volume} with the measured size of the extended emission from Table~\ref{tab:extendedprops}, the spectral fit parameters from Table~\ref{tab:spectralfits}, the Gaia the distance, and all the associated errors, we calculated the emission measure ($EM$), volume ($V$), and electron number density ($n_{\rm e}$) for the 2009 and 2011 observations and listed these values in Table~\ref{tab:spectralfits}. 
The volume appears to have increased $\sim30\%$, which is within the error range of the apparent decrease in the X-ray flux.
The $n_{\rm e}$ values suggest that the density has decreased from 2009 to 2011 by more than a factor of $\sim2$. 
The $n_{\rm e}$ values indicate cooling times ($t_{\rm cool}$) in excess of $10^{4} {\rm ~yr}$. 

\subsection{Mass of the Extended X-ray Emitting Material}

We estimated the mass of the X-ray emitting gas using $M_{\rm sh} \approx m_{\rm H} n_{\rm H} V$, giving,
\begin{equation}\label{eq:mass}
M_{\rm sh} \approx 1.7\times10^{-5} ~M_{\odot} \left(\frac{n_{\rm e}}{300 {\rm ~cm}^{-3}}\right) \left(\frac{V}{7\times10^{49} {\rm ~cm}^{3}}\right).
\end{equation}
The estimates for the 2009 and 2011 observations are provided in Table~\ref{tab:spectralfits} and are consistent within their respective error bars. 

Assuming the red giant is losing mass into a spherically symmetric shell at a rate of $\dot{M} \sim 10^{-7}$, we would expect the total mass reservoir of $\sim 2\times10^{-6}\ M_{\odot}$ provided by the red giant between the 1985 and 2006 eruptions. 
The spherical sectors we assume are $\sim 20\%$ of the entire spherical volume, a similar fraction of a spherically symmetric red giant wind would be swept up in collimated blast wave, suggesting that most of the X-ray emitting material may originate from the ejecta.  
\citet{2009ApJ...691..418D} find that a higher red giant mass loss rate of $\sim2\times10^{-6}\ M_{\odot}\ {\rm yr}^{-1}$ is needed, leading to a wind mass reservoir of $\sim 4\times10^{-5}\ M_{\odot}$, $20\%$ of which is $\sim8\times10^{-6}\ M_{\odot}$. 
Under these conditions, the majority of the detected X-ray emitting material could originate from the swept up red giant mass.  

\section{Conclusion} \label{sec:conclusions} 

A sequence of X-ray observations by the {\it Chandra} X-ray Observatory $\sim$3 and $\sim$5 years after the 2006 eruption of RS Oph reveal extended X-ray emission to the east and west of the binary system. 
The extended X-ray emission appears bipolar in nature with an opening angle of %$\sim
approximately $70^{\circ}$. 
The distance between the central binary and the most distant X-ray emitting gas to both the east and west increased from $1\farcs3$ in 2009 to $2\farcs0$ in 2011. 
This expansion rate is consistent with expansion seen in earlier radio and optical emission in a similar orientation. 
Overall, the multi-wavelength emission exhibits a projected expansion rate of $1.1\pm0.1 {\rm ~mas~day}^{-1}$, which, at Gaia parallax-derived distance of $2.4\pm0.3 {\rm ~kpc}$ and for a binary orbital inclination of $51\pm2^{\circ}$, suggests an expansion velocity of $\sim6000 {\rm ~km~s}^{-1}$. 

Spectral analysis of the X-ray emission provides no evidence for cooling of the X-ray emitting plasma, consistent with the free expansion of the material into the cavity excavated by the 1985 eruption.
The X-ray emission likely arises from a reverse shock driven into the bipolar flow and carried outward by the bipolar flow from the 2006 eruption. 
The orientation of the X-ray emission and the binary system suggests that equatorial enhancements play some role in collimating some ejecta away from the equatorial plane. 

The recurring nova eruptions from RS Oph provide an excellent astrophysical laboratory for studying nova eruptions and the shaping of their ejecta.
The latest eruption is likely driving ejecta into the cavity vacated by the 2006 eruption. 
Indeed, early multiwavelength observations after the most recent outburst also point to the presence of collimated structures in the same direction as those reported here \citep[e.g.][]{2021ATel14863....1N,2021ATel14860....1M}. 
The extended X-ray emitting material we identified, assuming free expansion, would have resided $\sim 6^{\prime\prime}$ from the binary system at the time of the latest eruption (2021  August  8.93  UT). 
As eruptions continue to drive material into these bipolar regions, it is possible that these relic shells may be detectable in future deep multiwavelength observations. 

\acknowledgments

Support for this work was provided by the National Aeronautics and Space Administration through Chandra Awards issued by the Chandra X-ray Center, which is operated by the Smithsonian Astrophysical Observatory for and on behalf of the National Aeronautics Space Administration under contract NAS8-03060. RMJ acknowledges additional support from NASA contract NAS8-03060. GJML is a member of the CIC-CONICET (Argentina) and acknowledge support from grants ANPCYT-PICT 0901/2017.

This research has made use of data obtained from the Chandra Data Archive and the Chandra Source Catalog, and software provided by the Chandra X-ray Center (CXC) in the application packages CIAO, ChIPS, and Sherpa. 
This research made use of Astropy, a community-developed core Python package for Astronomy \citep{2018AJ....156..123A, 2013A&A...558A..33A}. 
\vspace{5mm}
\facilities{CXO}

\software{CIAO \citep{2006SPIE.6270E..60F}, ChaRT \citep{2003ASPC..295..477C}, MARX \citep{2012SPIE.8443E..1AD}, \texttt{emcee} \citep{2013PASP..125..306F}, \texttt{astropy} \citep{2013A&A...558A..33A,2018AJ....156..123A}}

\end{document}